\documentclass[%
 aip,
 amsmath,amssymb,
 reprint,%
]{revtex4-1}

\usepackage{graphicx}% Include figure files
\usepackage{dcolumn}% Align table columns on decimal point
\usepackage{bm}% bold math
\usepackage[utf8]{inputenc}
\usepackage[T1]{fontenc}
\usepackage{mathptmx}
\usepackage{etoolbox}

\def\@email#1#2{%
 \endgroup
 \patchcmd{\titleblock@produce}
  {\frontmatter@RRAPformat}
  {\frontmatter@RRAPformat{\produce@RRAP{*#1\href{mailto:#2}{#2}}}\frontmatter@RRAPformat}
  {}{}
}%
\makeatother
\begin{document}

\preprint{AIP/123-QED}

\title[]{Phase Transitions and Virtual Exceptional Points in Quantum Emitters Coupled to Dissipative Baths}
\author{Stefano Longhi}
 \email{stefano.longhi@polimi.it}
 \altaffiliation[Also at ]{IFISC (UIB-CSIC), Instituto de Fisica Interdisciplinar y Sistemas Complejos - Palma de Mallorca, Spain}%Lines break automatically or can be forced with \\
\affiliation{ Dipartimento di Fisica, Politecnico di Milano, Piazza L. da Vinci 32, I-20133 Milano, Italy}
%\\This line break forced with \textbackslash\textbackslash

\date{\today}% It is always \today, today,
\begin{abstract}
Controlling atom-photon interactions in engineered environments is central to quantum optics and emerging quantum technologies. Non-Hermitian (NH) photonic baths, where dissipation fundamentally reshapes spectral and dynamical properties, provide versatile platforms for such control. Here we investigate the relaxation dynamics of a single two-level quantum emitter coupled to the edge of a semi-infinite dissipative bosonic lattice with uniform loss. Despite the simplicity of this bath, we uncover rich dynamical phase transitions, i.e. qualitative changes in spontaneous emission decay as system parameters are varied. In particular, we establish the existence of an optimal dissipative environment for accelerated spontaneous emission. The phase transitions are traced to spectral restructuring of the resolvent, in some cases governed by the coalescence of resonance states on the second Riemann sheet. We identify these coalescences as virtual exceptional points (EPs) of resonance origin, providing a conceptual bridge with EP physics while highlighting distinctive features of infinite-dimensional NH systems. More broadly, our results illustrate how the specific nature of dissipation -- whether uniform losses, staggered losses, or dephasing -- can profoundly impact emitter relaxation, pointing to dissipation engineering as a versatile tool for quantum technologies.
\end{abstract}

\maketitle

\section{Introduction}

Controlling atom-photon interactions at the quantum level is central to quantum optics and a key enabler of emerging quantum technologies \cite{T0,T1,T2,T3,T4}. Engineered photonic environments allow one to tailor spontaneous emission, stabilize bound states, and mediate long-range interactions, thereby providing versatile resources for quantum networks and light-matter interfaces. Recent advances in integrated and nanophotonics \cite{T3,T5}, superconducting circuits \cite{T2,T6,T7,T8}, and cold-atom arrays \cite{T9,T10} have significantly expanded the range of structured and dissipative baths available for such purposes.

Non-Hermitian (NH) physics \cite{T11,T11b,T12,T13,T14,T15,T15b} provides a powerful framework to describe these environments, where dissipation and loss fundamentally reshape spectral and dynamical properties. Recent works have addressed the dynamics of quantum emitters coupled to NH baths \cite{T16,T17,T18,T19,T20,T20b,T21}, predicting exotic effects such as unconventional emission dynamics \cite{T17}, hidden bound states of skin-effect origin \cite{T16,T18,T19}, non-reciprocal interactions between emitters \cite{T17,T20b}, algebraic atomic decay in lattices with staggered dissipation \cite{T18,T19}, and unusual in-gap chiral or extended photon-emitter dressed states \cite{T21}.

In finite-dimensional NH systems, a hallmark phenomenon is the appearance of exceptional points (EPs), where pairs of eigenvalues and eigenvectors coalesce \cite{EP1,EP2,EP3,EP4,EP5,EP5b}. EPs underpin striking dynamical effects and have been widely explored in both classical and quantum settings \cite{EP6,EP7,EP8,EP8b,EP9,EP9c,EP9d,EP9e,EP10,EP11,EP12,EP12a,EP12b,EP13,EP14,EP9b,EP15,EP16,EP17,EP18,EP18a,EP18b,EP18c,EP19,EP20}, with applications ranging from ultrasensitive sensing \cite{EP7,EP8,EP8b,EP9,EP10,EP11,EP12}, to topological and chiral mode transfer \cite{EP12a,EP12b,EP13,EP14,EP9b}, structured light generation \cite{EP15}, entanglement phenomena \cite{EP19}, and photon blockade \cite{EP18b}. Such studies have established EPs as a cornerstone of NH physics. Moreover, effective NH descriptions of EPs can be microscopically grounded by embedding them into open quantum system models \cite{GA-1,GA0,GA0b,GA1,GA1b,GA2,GA3,GA4,GA4b,GA5}, clarifying their physical origin and connection to system-bath dynamics. Importantly, however, in most of these works \cite{GA2,GA3,GA4b,GA5}  the bath itself is taken to be Hermitian, with non-Hermiticity introduced only through localized loss or gain terms acting on the system. By contrast, as in Refs.\cite{T16,T17,T18,T19,T20} here we focus on the complementary situation where the bath itself is intrinsically non-Hermitian (dissipative), a setting that leads to qualitatively different spectral and dynamical features. Beyond conventional EPs arising in effective NH descriptions of the underlying dissipative dynamics, several extensions have been introduced, including Liouvillian EPs in Markovian systems \cite{L1,L2,L3,L4,L5,L6}, virtual EPs in parametrically driven systems \cite{V1,V2}, and non-Markovian EPs in structured reservoirs \cite{NM1}, thereby broadening the scope of NH singularities.

In infinite-dimensional NH systems, the spectral structure is richer: in addition to discrete bound states and the continuous spectrum, resonance states can emerge as poles of the resolvent on the second Riemann sheet \cite{T11b,GA2,GA4,GA4b,GA5}.  In the present dissipative bath, the atom-photon bound states are spatially localized but decay exponentially in time owing to the uniform losses of the photonic lattice. While resonance states do not belong to the Hilbert space and are obtained via Siegert boundary conditions \cite{GA-1,GA1b}, they are well known to shape temporal relaxation dynamics \cite{GA-1,D1,D2,D3,D4,D4b,D5,D5b} and scattering features \cite{T11b,GA-1,GA0,GA0b}, such as Breit--Wigner or Fano resonances, in infinite-dimensional systems. 

In this work, we show that the coalescence of atom-photon bound states and resonance states, through physical or virtual EPs, can induce dynamical phase transitions \cite{Teza,Longhi} in the spontaneous emission process of a quantum emitter coupled to a dissipative bosonic bath. A key consequence of these transitions is the emergence of an optimal dissipation strength that maximizes the spontaneous emission rate of the emitter. Too little or too much dissipation slows down relaxation, while an intermediate value yields the fastest decay. Some of these transitions are associated with resonance coalescence -- conceptually related to virtual EPs of resonance origin \cite{GA2} -- and are therefore distinct from discrete EP degeneracies of NH Hamiltonians, Liouvillian EPs \cite{L1,L2}, or spectral singularities in the continuous spectrum of infinite-dimensional NH systems \cite{SS1,SS2,SS3,SS4}. 
We illustrate these ideas in a minimal dissipative model of relevance to waveguide QED: a quantum emitter coupled to the edge of a semi-infinite bosonic lattice with uniform dissipation. We uncover three distinct coupling regimes -- weak, intermediate, and strong -- where different types of phase transitions arise from the restructuring of poles of the resolvent on the first and second Riemann sheets. Unlike standard EP-related effects, some transitions originate from spectral restructuring due to resonance coalescence, i.e., the emergence of resonance-driven virtual EPs. Our results establish dissipative baths as simple yet powerful platforms for exploring NH phase transitions and optimal relaxation dynamics. Uniformly lossy lattices, being both analytically tractable and experimentally accessible in nanophotonic waveguides, superconducting circuits, and cold-atom setups, provide ideal testbeds. Beyond their fundamental significance, these findings suggest strategies for harnessing dissipation to accelerate relaxation and enhance atom-photon coupling, with extensions to multi-emitter and higher-dimensional baths expected to further connect these ideas with cutting-edge experimental platforms.

\section{Photon emission in a dissipative bosonic lattice: model and basic equations}  

\begin{figure}[h!]
\centering
\includegraphics[width=0.8\linewidth]{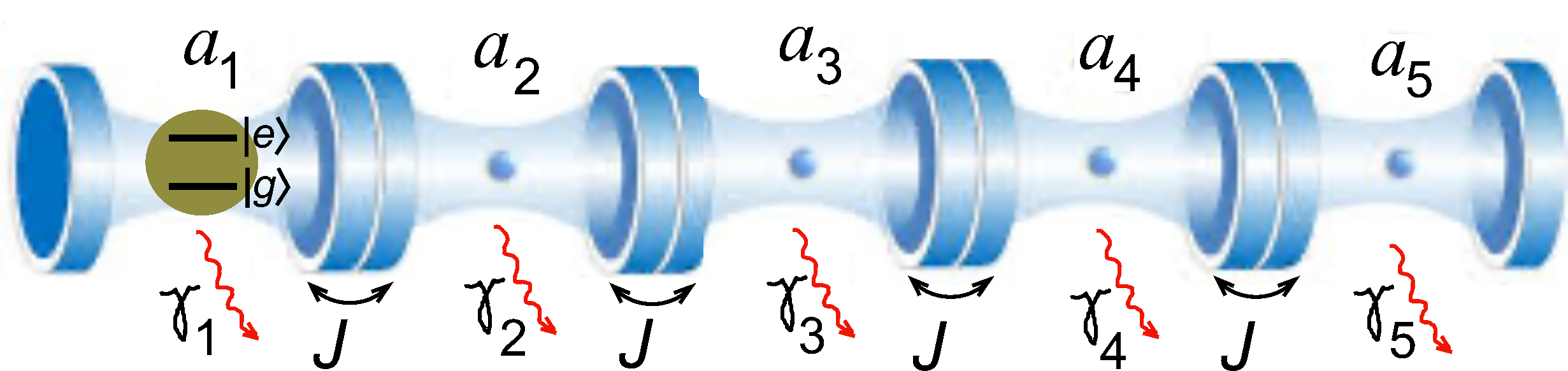} 
\caption{Schematic of a two-level quantum emitter coupled to a dissipative bosonic bath consisting of a semi-infinite array of coupled optical cavities. The emitter is placed inside the edge resonator ($n=1$). Here, $J$ denotes the photon hopping rate between adjacent resonators, $g_0$ the atom-photon coupling strength, and $\gamma_n$ the photon loss rates in each resonator.}
\label{fig:Fig1}
\end{figure}

We consider a standard model in waveguide QED, where a quantum emitter, modeled as a two-level atom, is coupled to a nanophotonic lattice of resonators \cite{T17,T18,T19,MD1,MD2,MD3,MD4,MD5,MD6,MD7}, as illustrated in Fig.~\ref{fig:Fig1}. Photon modes in the lattice resonators experience local loss, rendering the bosonic bath non-Hermitian \cite{T17,T18,T19,MD8}.  

Let $|e\rangle$ and $|g\rangle$ denote the excited and ground states of the atom, with transition frequency $\omega_0$, placed inside the edge resonator $n=1$ of the semi-infinite array, and let $\omega_c \simeq \omega_0$ be the resonance frequency of the photon modes. Under the Markovian and rotating-wave approximations, the time evolution of the atom-photon density operator $\rho(t)$ is governed by the Lindblad master equation ($\hbar=1$) \cite{T18}:  
\begin{equation}
\frac{d \rho}{dt}=-i [H, \rho]+  \sum_{n=1}^{\infty} \gamma_n \left(  2 a_n \rho a_n^{\dag}-a^{\dag}_n a_n \rho-\rho a_n^{\dag} a_n \right) \equiv \mathcal{L} \rho,
\label{eq:master}
\end{equation}
where  
\begin{eqnarray}
H & = & (\omega_0-\omega_c)  |e \rangle \langle e|- \sum_{n=1}^{\infty}  \left\{J(a^{\dag}_{n+1}a_n+{\rm H.c.}) \right\} \nonumber \\
&+ & g_0 \left( a_1^{\dag} |g \rangle \langle e| +{\rm H.c.} \right),
\label{eq:H}
\end{eqnarray}
is the full atom-photon Hamiltonian in the rotating-wave approximation. Here, $\gamma_n$ is the photon loss rate in the $n$-th resonator, $J$ is the hopping rate between adjacent resonators, $a^{\dag}_n$ ($a_n$) is the photon creation (annihilation) operator in the $n$-th resonator, and $g_0$ is the atom-photon coupling strength.  

Previous studies considered staggered losses corresponding to passive parity-time symmetry of the NH bath \cite{T18}, leading to robust algebraic atomic decay. Here, we focus on the simpler case of uniform dissipation, $\gamma_n = \gamma$. Despite the simplicity of the bath, the model exhibits dynamical phase transitions, some of which are driven by the coalescence of resonance states, identified as \emph{virtual exceptional points}.  

To describe spontaneous emission, we assume that at $t=0$ the atom is in the excited state and the photon field is in the vacuum state, i.e. $\rho_0 = \rho(t=0) = |\psi_0 \rangle \langle \psi_0|$, with $|\psi_0\rangle = |e\rangle \otimes |0\rangle$, where $|0\rangle$ is the photon vacuum state. Since the system has no gain, the dynamics is confined to the $N \leq 1$ excitation sector of Hilbert space, and quantum jump terms in the Lindblad master equation (1) do not affect the emitter's relaxation  \cite{T18}. Accordingly:
\begin{eqnarray}
\rho(t) & = & e^{-i H_{NH} t} \rho_0 e^{i H_{NH}^{\dag} t} + p_t |g \rangle \langle g| \otimes |0 \rangle \langle 0| \nonumber \\
& = & |\psi(t)\rangle \langle \psi(t)| + p_t |g \rangle \langle g| \otimes |0 \rangle \langle 0|,
\label{eq:rho}
\end{eqnarray}
where the effective NH Hamiltonian is  
\begin{eqnarray}
H_{NH} & = & H - i \gamma \sum_{n=1}^{\infty} a_n^\dag a_n \nonumber \\
& = & \Delta \omega_0 |e\rangle\langle e| - \sum_{n=1}^{\infty} \left\{ J(a_{n+1}^\dag a_n + {\rm H.c.}) \right\} \nonumber \\
&+ & g_0 \left( a_1^\dag |g\rangle \langle e| + {\rm H.c.} \right) - i \gamma \sum_{n=1}^{\infty} a_n^\dag a_n,
\label{eq:HNH}
\end{eqnarray}
with $\Delta \omega_0 = \omega_0 - \omega_c$, and  
\begin{equation}
|\psi(t)\rangle = e^{-i H_{NH} t} |\psi_0\rangle, \quad 
p_t = 1 - {\rm Tr} \left( e^{-i H_{NH} t} \rho_0 e^{i H_{NH}^\dag t} \right).
\label{eq:pt}
\end{equation}
This means that, as quantum jumps do not play any role in the single excitation sector, the dynamics is fully captured by the quantum state $| \psi(t) \rangle$ that evolves according to the effective NH Hamiltonian $H_{NH}$ \cite{T18,Roccati}. Writing  
\begin{equation}
|\psi(t)\rangle = c_a(t) |e\rangle \otimes |0\rangle + \sum_{n=1}^{\infty} b_n(t) |g\rangle \otimes a_n^\dag |0\rangle,
\label{eq:psi_t}
\end{equation}
the Schr\"odinger equation $i \partial_t |\psi(t)\rangle = H_{NH} |\psi(t)\rangle$ yields
\begin{eqnarray}
i \frac{dc_a}{dt} & = & \Delta \omega_0 c_a + g_0 b_1, \nonumber \\
i \frac{db_1}{dt} & = & g_0 c_a - J b_2 - i \gamma b_1, \\
i \frac{db_n}{dt} & = & -J (b_{n+1} + b_{n-1}) - i \gamma b_n, \quad n \ge 2. \nonumber
\label{eq:coupled}
\end{eqnarray}
These equations can be solved exactly using standard spectral methods \cite{GA2,SS3,Y1,Y2}. Introducing  
\begin{equation}
c(k,t) = \sqrt{\frac{2}{\pi}} \sum_{n=1}^{\infty} b_n(t) \sin(nk), \quad 0 \le k < \pi,
\label{eq:c_k}
\end{equation}
the dynamics reduces to a non-Hermitian Friedrichs-Lee (Fano-Anderson) model \cite{SS3,EPJB}:
\begin{eqnarray}
i \frac{dc_a}{dt} & = & \Delta \omega_0 c_a + \int_0^\pi dk g(k) c(k,t) , \label{eq:FA1} \\
i \frac{dc(k,t)}{dt} & = & \omega(k) c(k,t) + g^*(k) c_a, \label{eq:FA2}
\end{eqnarray}
with  
\begin{equation}
\omega(k) = -2 J \cos k - i \gamma, \quad g(k) = \sqrt{\frac{2}{\pi}} g_0 \sin k, 
\label{eq:dispersion}
\end{equation}
where $\omega(k)$ is the NH photonic bath dispersion and $g(k)$ is the atom-photon spectral coupling. The survival probability is  
\begin{equation}
P_s(t) = {\rm Tr}(\rho(t) \rho_0) = |c_a(t)|^2.
\label{eq:Ps}
\end{equation}
Due to the non-Hermitian, infinite-dimensional nature of $H_{NH}$, in addition to conventional (or physical) EPs -- coalescence of eigenenergies and eigenstates in the point spectrum-- \emph{virtual EPs} can emerge from the coalescence of resonant states. A detailed discussion of the difference between ordinary (or physical) EPs and virtual EPs is provided in the Appendix A.

\begin{figure}[h!]
\centering
\includegraphics[width=0.8\linewidth]{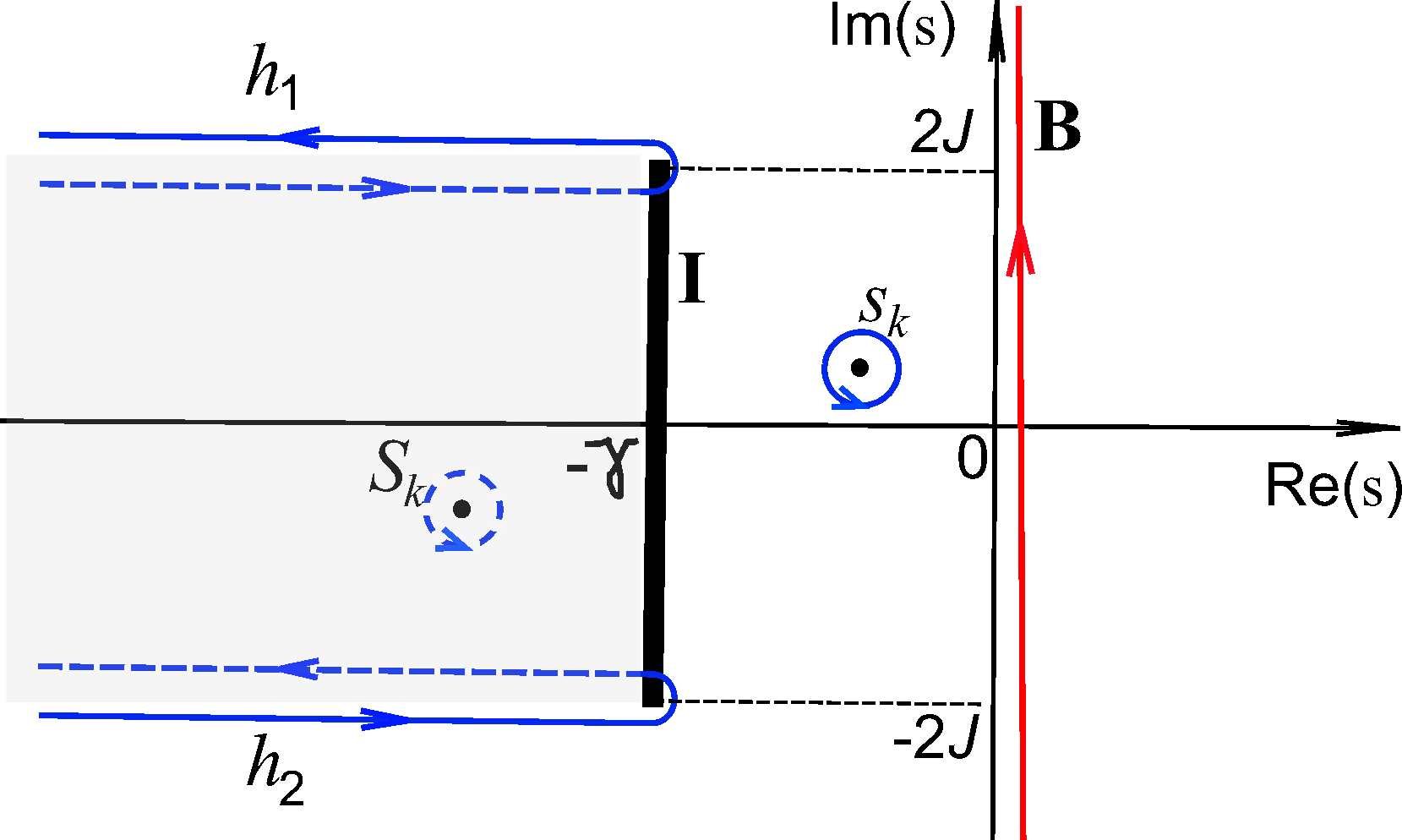}  
\caption{Contour paths in the complex $s$-plane for Eq.~\eqref{eq:Laplace}. The solid segment $\mathbf{I}$ along ${\rm Re}(s)=-\gamma$ is the branch cut of the self-energy $\Sigma(s)$. The Bromwich path $\mathbf{B}$ can be deformed into the Hankel paths $h_1$ and $h_2$, and the pole contributions $s_k$ (first Riemann sheet, bound states) and $S_k$ (second Riemann sheet, resonant states). The shaded region indicates the domain of analytic continuation for $\hat{c}_a(s)$ on the second Riemann sheet.}
\label{fig:Fig2}
\end{figure}

\section{Relaxation dynamics, virtual exceptional points, and dynamical phase transitions}  

\subsection{Relaxation dynamics: general}  

The exact solution for $c_a(t)$, governed by Eqs.~\eqref{eq:FA1}-\eqref{eq:FA2}, can be obtained via Laplace or resolvent methods \cite{T18,D4b,D5,SS3,Y1,Y2}:

\begin{equation}
c_a(t) = \frac{1}{2\pi i} \int_{\rm B} ds \, \hat{c}_a(s) e^{st}, \quad 
\hat{c}_a(s) = \frac{1}{s + i \Delta \omega_0 + i \Sigma(s)},
\label{eq:Laplace}
\end{equation}
with the self-energy  
\begin{equation}
\Sigma(s) = \int_0^\pi dk \frac{|g(k)|^2}{is - \omega(k)} = i \frac{g_0^2}{2 J^2} \left( s + \gamma - \sqrt{4J^2 + (s+\gamma)^2} \right), \label{self}
\end{equation}
and Bromwich path ${\rm B} = (-i\infty + 0^+, i\infty + 0^+)$. The Laplace transform $\hat{c}_a(s)$ is analytic in the entire complex $s$ plane, except for a  branch cut ${\bf I}$ along the segment $s \in (-2 i J - \gamma, 2 i J - \gamma)$ and possible poles on the semi-half complex plane ${\rm Re}(s) \ge -\gamma$. The branch cut
corresponds to the absolutely continuous spectrum of $H_{NH}$, whereas the poles of $\hat{c}_a(s)$ correspond to the point spectrum (atom-photon bound states). The integration Bromwich path B can be deformed as illustrated in Fig.2.
Deforming the contour requires to cross the branch cut I,  from the right to the left side. Therefore, analytic continuation $\hat{c}_a^{(II)}(s)$ of $\hat{c}_a(s)$ on the second Riemann sheet, obtained from Eq.(13) by replacing the self-energy with its analytic continuation
\begin{equation}
\Sigma^{(II)}(s) = \Sigma(s) + i \frac{g_0^2}{J^2} \sqrt{(s+\gamma)^2 + 4J^2}
\end{equation}
should be considered in the shaded area of Fig.2.
Hence, $c_a(t)$ has contributions from:  
(i) poles $s_k$ of $\hat{c_a}(s)$ on first Riemann sheet (bound states),  
(ii) poles $S_k$ of $\hat{c}_a^{(II)}(s)$ on the second Riemann sheet (resonant states), and  
(iii) Hankel path integrals $H_1(t), H_2(t)$ along the contours $h_{1}$, $h_2$ in the first and second Riemann sheets, i.e.
\begin{equation}
c_a(t) = \sum_k r_k e^{s_k t} + \sum_k R_k e^{S_k t} + H_1(t) + H_2(t),
\end{equation}
with  
\begin{eqnarray}
H_1(t) & = & C_1(t) e^{-\gamma t + 2 i J t}, \quad
H_2(t) = C_2(t) e^{-\gamma t - 2 i J t}, \\
C_1(t) & = & \int_0^\infty dx \, [ f_+^{(II)}(x) - f_+(x) ] e^{-xt}, \\
C_2(t) & = &  \int_0^\infty dx \, [ f_-(x) - f_-^{(II)}(x) ] e^{-xt}.
\end{eqnarray}
In the above equations,  we have set 
$f_\pm(x) = \hat{c}_a(s=-\gamma-x \pm 2 i J)$ and $f_\pm^{(II)}(x) = \hat{c}_a^{(II)}(s=-\gamma-x \pm 2 i J)$, $r_{k}$ and  $R_k$ are the residues of $\hat{c}_a(s)$ and  $\hat{c}_a^{(II)}(s)$ at the poles $s_k$ and $S_k$  on the first and second Riemann sheets, given by
\[
r_k=\frac{s_k (\sigma-1)+\gamma \sigma}{(2 \sigma-1)s_k+ \gamma \sigma} 
\]
\[ 
R_k=\frac{S_k (\sigma-1)+\gamma \sigma}{(2 \sigma-1)S_k+\gamma \sigma},
\]
 and $\sigma \equiv g_0^2/(2J^2)$ is the normalized atom-photon coupling strength. 
The Hankel path contributions produce non-exponential decay features at short and long times, with $C_{1,2}(t) \sim t^{-3/2}$ at long times \cite{Y2}.  
Poles of $\hat{c}_a(s)$ and $\hat{c}_a^{(II)}(s)$ govern the relaxation dynamics at intermediate and long-time scales.
 According to Eqs.~\eqref{eq:Laplace} and \eqref{self}, the poles on the first or second Riemann sheets satisfy the quadratic equation
\begin{eqnarray}
(1-2\sigma)s^2 - 2[\gamma \sigma + i \Delta\omega_0(\sigma-1)] s  \nonumber \\ 
- 4\sigma^2 J^2 - \Delta\omega_0^2 - 2 i \gamma \sigma \Delta \omega_0 = 0.
\label{eq:pole_eq}
\end{eqnarray}
 Only poles with ${\rm Re}(s) \le 0$ contribute to the dynamics. Since Eq.~(20) is quadratic, at most two poles---located either on the first or on the second Riemann sheet---can contribute to the decay dynamics of $c_a(t)$. A pole $s_k$ on the first Riemann sheet corresponds to an atom--photon bound state, in which case the condition $\mathrm{Re}(s_k) \geq -\gamma$ is satisfied. A pole on the second Riemann sheet instead corresponds to either a resonant or an anti-resonant state (see Appendix~B for technical details). For a resonant state one finds $\mathrm{Re}(S_k) < -\gamma$, i.e., the pole lies to the left of the branch cut. By contrast, anti-resonant states are always associated with a pole $S_k$ such that $\mathrm{Re}(S_k) > 0$; consequently, they do not contribute to the decay dynamics of $c_a(t)$.

As the parameters $g_0/J$, $\Delta \omega_0/J$, and $\gamma/J$ are varied, the poles move in the complex plane, leading to different relaxation behaviors. Coalescence of poles on the first Riemann sheet corresponds to ordinary EPs, while coalescence on the second Riemann sheet corresponds to virtual EPs (see Appendix A). From Eq.~\eqref{eq:pole_eq}, a necessary condition for pole coalescence $s_{p_1} = s_{p_2}$ is $\Delta \omega_0 = 0$, i.e., atom and photon fields in resonance.  
In this resonant case, the two poles are
\begin{equation}
s_{p_{1,2}} = \frac{\gamma \pm \sqrt{\gamma^2 + 4(J^2 - g_0^2)}}{2 \left( J^2/g_0^2 - 1 \right)}.
\label{eq:poles}
\end{equation}
The corresponding form of bound or resonant eigenstates can be directly obtained by solving the eigenvalue equation $H_{NH} |\psi\rangle = E |\psi\rangle$ on the semi-infinite lattice, as shown in Appendix B. In particular, the coalescence of the two poles, $s_{p_1}=s_{p_2}$, always corresponds to the coalescence of the corresponding eigenstates, i.e. to either a virtual (for resonant states) or a physical (for bound states) EP.

The general relaxation dynamics is then fully determined by the positions of these poles, the residues $r_k, R_k$, and the Hankel contributions $H_1(t), H_2(t)$, giving a rich interplay of exponential, non-exponential, and long-time algebraic decay behavior in the spontaneous emission process of the quantum emitter, as discussed  in the next subsection.

\subsection{Dissipation-induced exceptional points and dynamical phase transitions}
The position of the poles $s_{p_1}$ and $s_{p_2}$ in the complex energy plane, and thus the appearance of EPs, critically depends on the normalized atom-photon coupling constant $g_0/J$. Specifically, three distinct cases should be considered: the weak-coupling regime ($g_0/J<1$), the moderate-coupling regime ($1<g_0/J< \sqrt{2}$), and the strong-coupling regime ($g_0/J > \sqrt{2}$. 

\subsubsection{Weak-coupling regime}
The weak coupling regime corresponds to the condition $g_0 / J <1$. In this regime, the two poles $s_{p_1}$ and $s_{p_2}$ are located on the real $s$ axis and never cross as the bath dissipation rate $\gamma$ is varied, i.e. there are not EPs of any kind. A typical behavior of the real and imaginary parts of the two poles $s_{p_1}$ and $s_{p_2}$ versus normalized loss rate $\gamma /J$ is shown in Figs.3(a) and (b). For $\gamma< \gamma_{c_1}$, with
\begin{equation}
\gamma_{c_1} = \frac{g_0^2}{J}
\end{equation}
both poles are on the second Riemann sheet and correspond to a resonant state (the pole $s_{p_1}$ with negative real part, smaller than $-\gamma$) and to an anti-resonant state (the pole $s_{p_2}$ with positive real part). The latter does not play any role in the relaxation dynamics and can be thus disregarded. In this case, the decay law (16) is specialized as [Fig.4(a)]
\begin{equation}
c_a(t)=R_1 \exp(S_1 t)+H_1(t)+H_2(t)
\end{equation}
\begin{figure}
\includegraphics[width=8.5 cm]{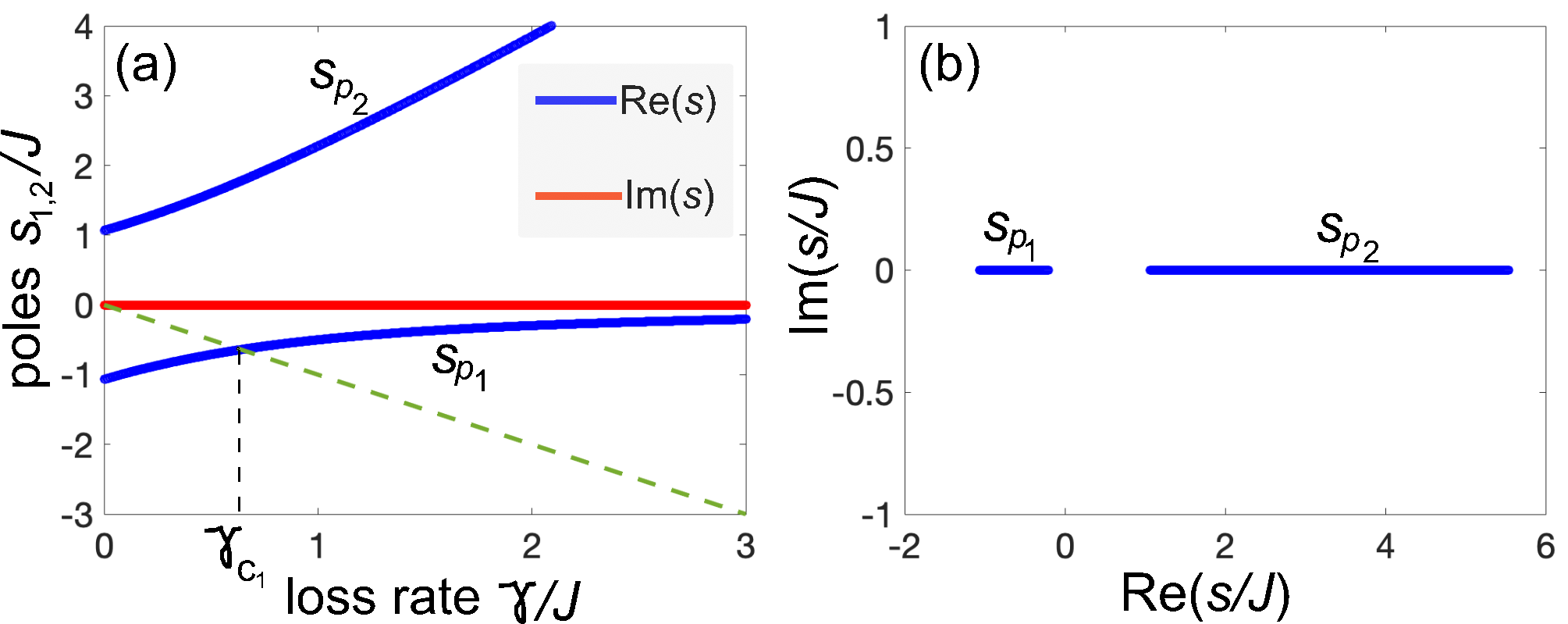}  
\caption{(a) Behavior of the real and imaginary parts of the two poles $s_{p_1}$ and $s_{p_2}$ versus loss rate $\gamma$ in the weak coupling regime ($g_0/J=0.8$). The dashed curve, corresponding to ${\rm Re}(s)=-\gamma$, separates the two Riemann sheets. Pole $s_{p_2}$, with positive real part, is located on the second Riemann sheet and does not impact on the relaxation dynamics. Pole $s_{p_1}$ is located on the second Riemann sheet for $\gamma < \gamma_{c_1} \equiv g_0^2/J=0.64 J$, corresponding to a resonant state, whereas it is located on the first Riemann sheet for $\gamma> \gamma_{c_1}$, corresponding to an atom-photon bound state. (b) Loci of the two poles $s_{p_{1,2}}$ in complex $s$ plane as the loss rate $\gamma$ is varied from $\gamma=0$ to $\gamma=3J$.}
\end{figure}
where $S_1=s_{p_1}$. An example of the relaxation dynamics is shown in Fig.4(b). After an initial short time (Zeno region), in the intermediate time region  the dominant contribution to the dynamics is given by the pole contribution on the second Riemann sheet, with a decay rate equal to $| {\rm Re}(s_{p_1})|> \gamma$. However, at longer times the dominant contribution comes from the Hankel paths $H_{1,2}(t)$ [Eqs.(17-19)], whose interference yields an oscillatory decay behavior at the characteristic frequency $4J$ enveloped by the damping function $\sim t^{-3/2} \exp(-\gamma t)$, as clearly shown in Fig.4(b).  Therefore, in this regime the long-time decay behavior deviates from an exponential and displays damped oscillations due to interference effects.\\
For $\gamma > \gamma_{c_1}$, the pole $s_{p_1}$ crosses the branch cut and is now located on the first Riemann sheet [Fig.5(b)]. Thus one has
\begin{equation}
c_a(t)=r_1 \exp(s_1 t)+H_1(t)+H_2(t)
\end{equation}
where $s_1=s_{p_1}$. In the temporal domain, the branch crossing of pole $s_1$ corresponds to a phase transition in the relaxation dynamics \cite{Teza,Longhi}, i.e. the asymptotic time behavior of $c_a(t)$ toward zero is not anymore dominated by the Hankel path contributions $H_{1,2}(t)$, as in Fig.4(b), rather it is purely exponential and dominated by the pole contribution corresponding to the atom-photon bound state, as shown in Fig.5(b).\\
Finally, an inspection of Fig.3(a) indicates that the spontaneous emission process of the quantum emitter in the asymptotic long-time regime is fastest at the critical loss rate $\gamma=\gamma_{c_1}=g_0^2/J$, with the largest decay rate given by $|s_{p_1}|=\gamma_{c_1}$. In fact, for $\gamma < \gamma_{c_1}$ the asymptotic relaxation dynamics is dominated by the Hankel path contributions, with a decay $c_a(t) \sim t^{-3/2} \exp(-\gamma t)$  that gets faster as $\gamma$ is increased. However, as $\gamma$ is increased above the critical value $\gamma_{c_1}$ the dominant decay rate is given by the atom-photon bound state contribution, which decreases as $\gamma$ is increased and vanishes as $\gamma / J \rightarrow \infty$. Physically, in this limit the dissipation in the bath is so large that the quantum emitter becomes decoupled to it as a result of an effective quantum Zeno dynamics \cite{Zeno}.

\begin{figure}
\includegraphics[width=8.5 cm]{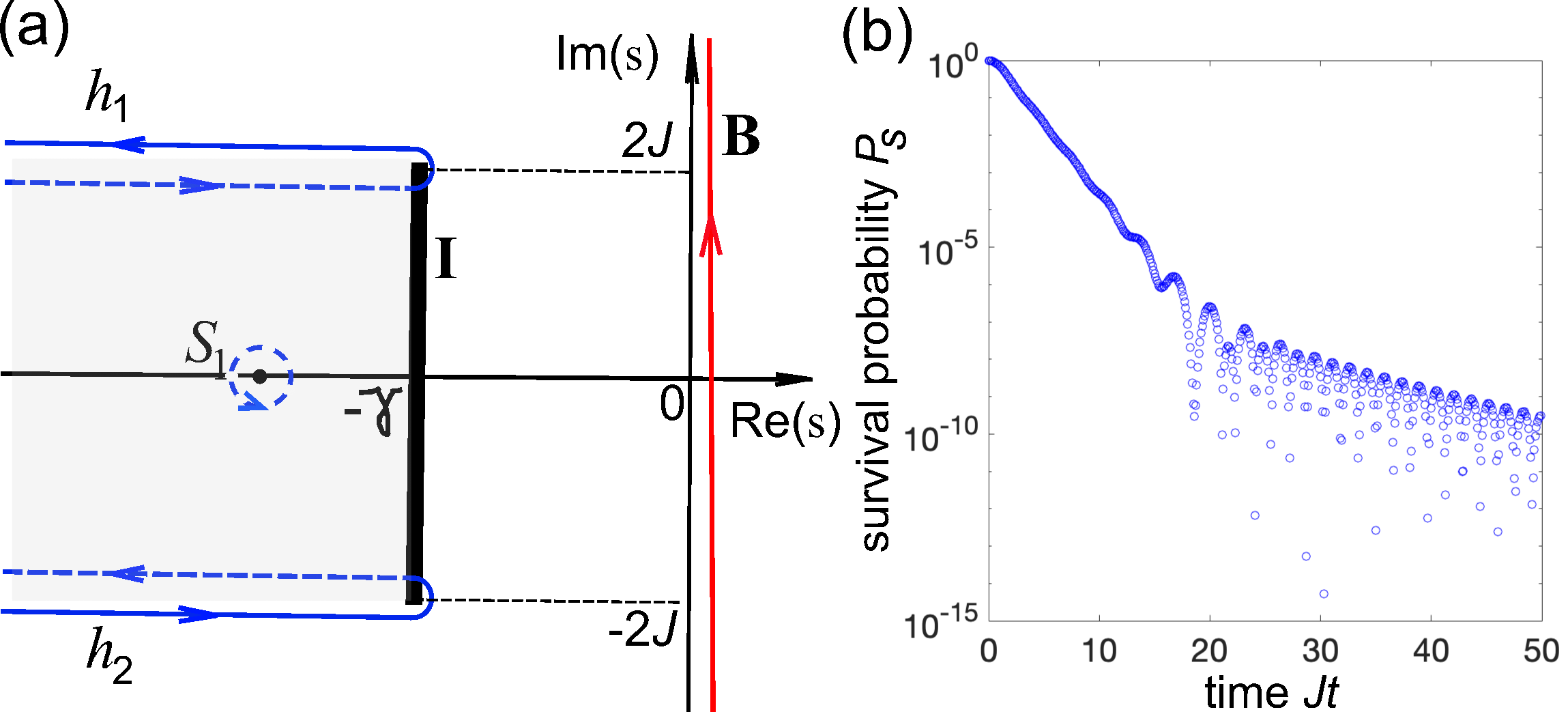}  
\caption{(a) Integration contour in complex $s$ plane in the weak coupling regime and for $\gamma< \gamma_{c_1}$. The contribution to the decay law $c_a(t)$ comes from the two Hankel paths $h_{1,2}$ and from the pole $S_1=s_{p_1}$ on the second Riemann sheet (resonant state). (b) Numerically computed behavior of the survival probability $P_s(t)=|c_a(t)|^2$ versus normalized time $Jt$ for parameter values $g_0/J=0.6$ and $\gamma/J=0.05$. }
\end{figure}
\begin{figure}
\includegraphics[width=8.5 cm]{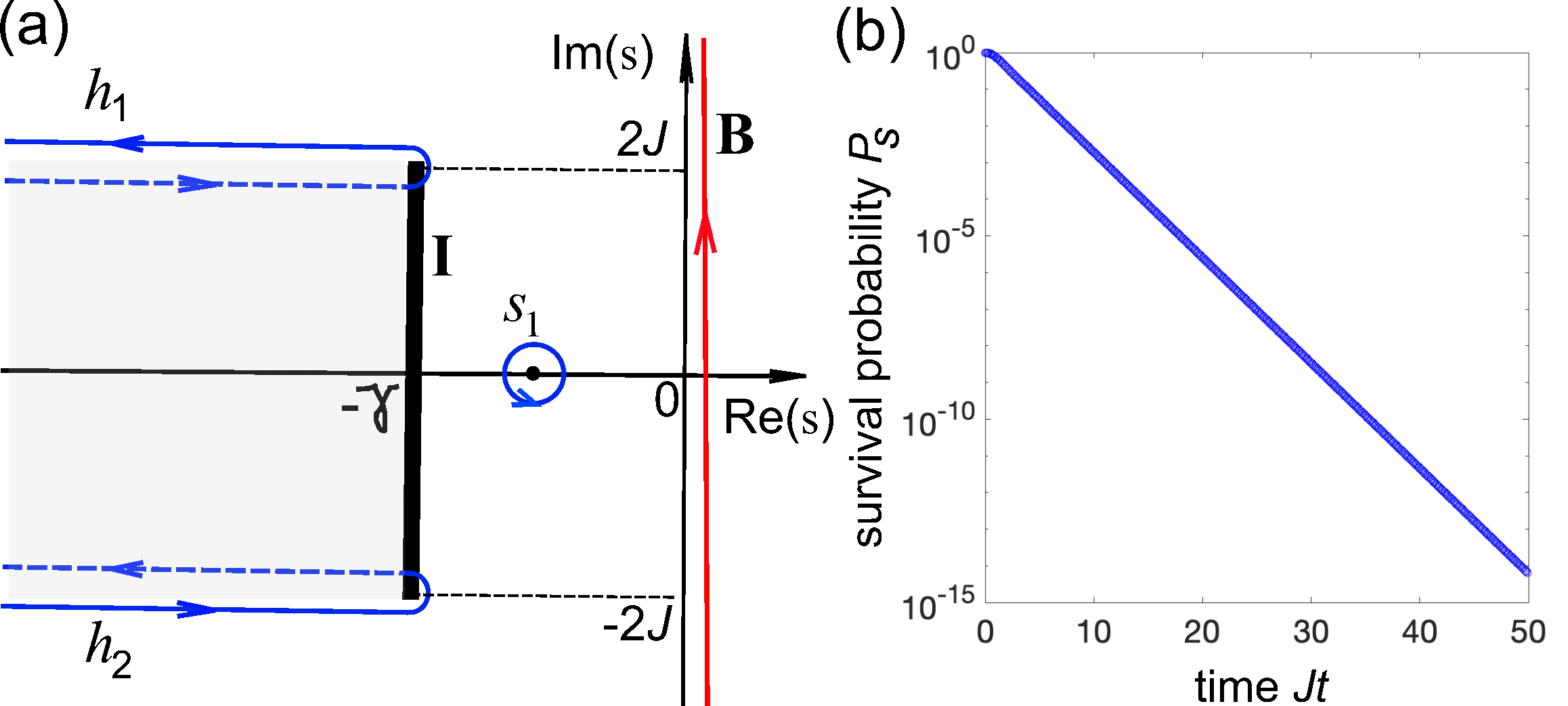}  
\caption{ (a) Integration contour in complex $s$ plane in the weak coupling regime and for $\gamma> \gamma_{c_1}$. (b) Numerically computed behavior of the survival probability $P_s(t)=|c_a(t)|^2$ versus normalized time $Jt$ for parameter values $g_0/J=0.6$ and $\gamma/J=0.5$. Note that the pole $s_{p_1}$ has now crossed the branch cut and is located on the first Riemann sheet. The contribution to the decay law $c_a(t)$ comes from the two Hankel paths $h_{1,2}$ and from the pole $s_1=s_{p_1}$ (bound state). Note the different long-time relaxation behavior of the survival probability in (b) as compared to Fig.4(b).}
\end{figure}

\subsubsection{Moderate-coupling regime}
The moderate coupling regime corresponds to the condition $1<g_0 / J < \sqrt{2}$. A typical behavior of the real and imaginary parts of the two poles $s_{p_1}$ and $s_{p_2}$ versus normalized loss rate $\gamma /J$ is shown in Figs.6(a) and (b). As it can be seen, for $\gamma< \gamma_{c_1}=g_0^2/J$ the two poles $s_{p_1}$ and $s_{p_2}$ lie on the second Riemann sheet and correspond to resonant states.  For $g_0$ not too close to $J$, the two poles are embedded in the shaded region of Fig.2, between the two Hankel paths $h_1$ and $h_2$, and thus  the decay law (16) takes the form [Fig.7(a) and (c)]
\begin{equation}
c_a(t)=R_1 \exp(S_1 t)+R_2 \exp(S_2 t)+H_1(t)+H_2(t)
\end{equation}
\begin{figure}
\includegraphics[width=8.5 cm]{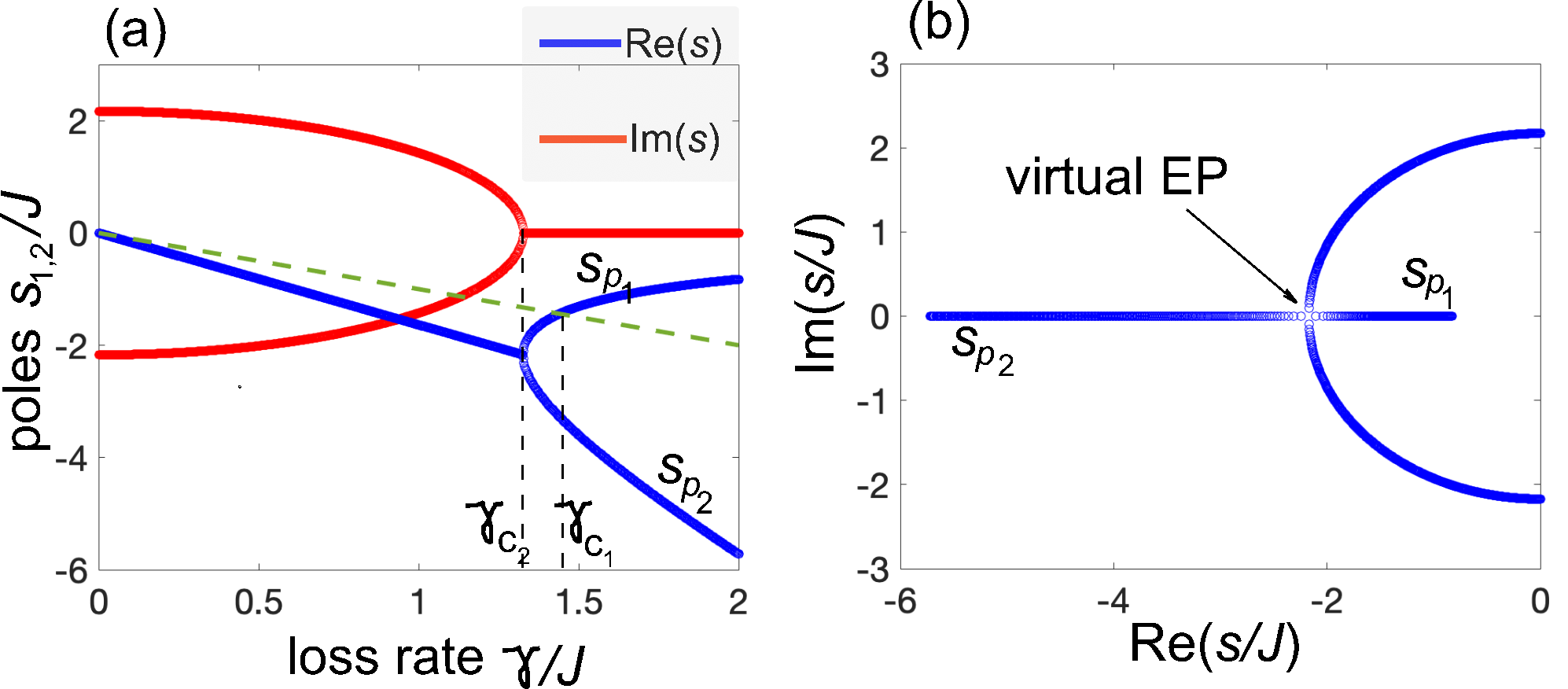}  
\caption{(a) Behavior of the real and imaginary parts of the two poles $s_{p_1}$ and $s_{p_2}$ versus loss rate $\gamma$ in the moderate coupling regime ($g_0/J=1.2$). The dashed curve, corresponding to ${\rm Re}(s)=-\gamma$, separates the two Riemann sheets. Both poles contribute to the relaxation dynamics. For $\gamma< \gamma_{c_1}=g_0^2/J$, the two poles lie on the second Riemann sheet and correspond to two resonances. At $\gamma=\gamma_{c_2}$, with $\gamma_{c_2}$ given by Eq.(27), the two poles coalesce, corresponding to a virtual EP. (b) Loci of the two poles in complex $s$ plane as the loss rate $\gamma$ is varied from $\gamma=0$ to $\gamma=2J$.}
\end{figure}
where $S_1=s_{p_1}$ and  $S_2=s_{p_2}$. The Hankel contributions are responsible for deviations from exponential decay at short and long time scales, and they dominate the asymptotic (long-time) relaxation dynamics since $| {\rm Re}( S_{1,2})|> \gamma$. Therefore, for $\gamma< \gamma_{c_1}$ the long-time relaxation dynamics is the same as in the weak coupling regime and displays a non-exponential oscillatory decay [Figs.7(b) and (d)] . For $\gamma> \gamma_{c_1}$ one of the two poles, $s_{p_1}$, crosses the branch cut and moves onto the first Riemann sheet, corresponding to the emergence of an atom-photon bound state [Fig.8(a)]. The decay law (16) takes then the form 
\begin{equation}
c_a(t)=r_1 \exp(s_1 t)+R_1 \exp(S_1 t)+H_1(t)+H_2(t)
\end{equation}
where $s_1=s_{p_1}$ and  $S_1=s_{p_2}$.  In this case the asymptotic long-time decay of $c_a(t)$ is exponential and dominated by the pole $s_{p_1}$ [Fig.8(b)], with a fastest decay attained at the critical value $\gamma=\gamma_{c_1}$.\\ 
An intriguing behavior, which is not found in the weak coupling regime, is the coalescence of the two poles $s_{p_1}$ and $s_{p_2}$ on the second Riemann sheet, corresponding to a virtual EP, at the critical value (Fig.6)
\begin{equation}
\gamma_{c_2}=2 \sqrt{g_0^2-J^2}.
\end{equation}

\begin{figure}
\includegraphics[width=8.5 cm]{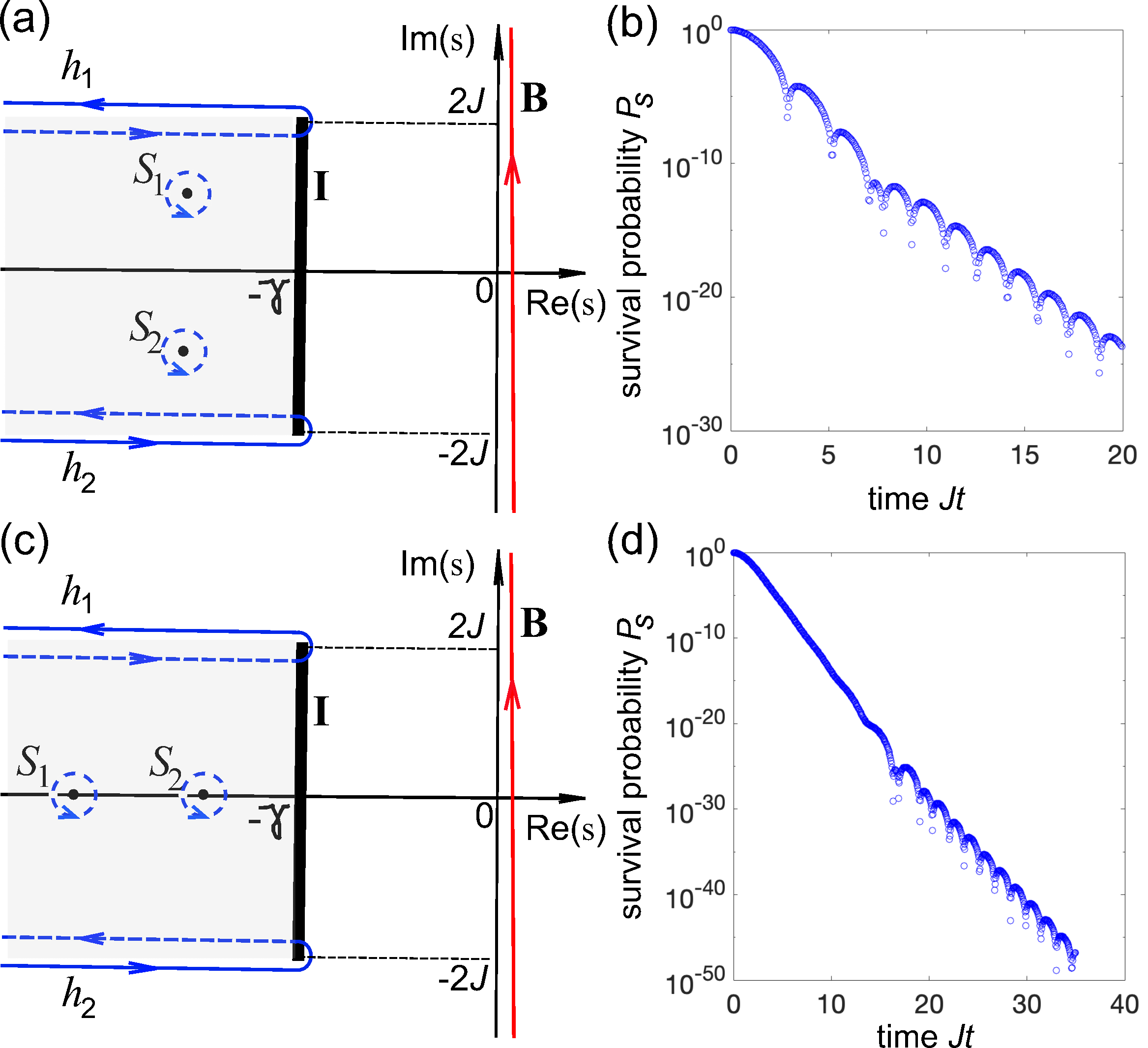}  
\caption{(a,c) Integration contour paths in complex $s$ plane in the moderate coupling regime for (a) $\gamma< \gamma_{c_2}$, and (c) $\gamma_{c_2}<\gamma<\gamma_{c_1}$. The contribution to the decay law $c_a(t)$ comes from the two Hankel paths $h_{1,2}$ and from the poles $S_1=s_{p_1}$ and $S_2=s_{p_2}$  on the second Riemann sheet (resonances). (b,d) Numerically computed behavior of the survival probability $P_s(t)=|c_a(t)|^2$ versus normalized time $Jt$ for $g_0/J=1.2$. In (b),  $\gamma/J=1,1$, corresponding to $\gamma< \gamma_{c_2}$; in (d), $\gamma /J=1.35$, corresponding to $\gamma_{c_2}<\gamma<\gamma_{c_1}$.}
\end{figure}

\begin{figure}
\includegraphics[width=8.5 cm]{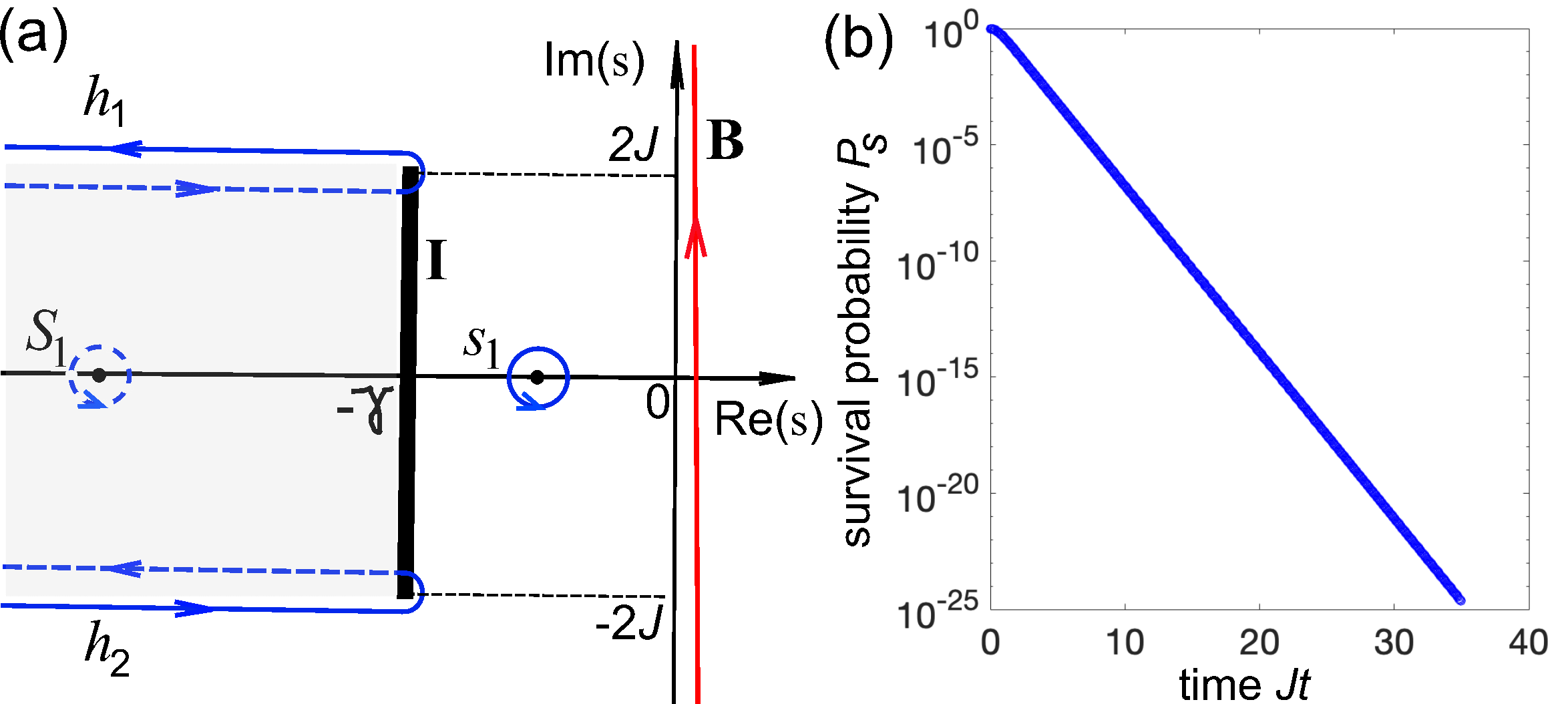}  
\caption{(a) Integration contour in complex $s$ plane in the moderate-coupling regime and for $\gamma> \gamma_{c_1}$. The contribution to the decay law $c_a(t)$ comes from the two Hankel paths $h_{1,2}$, from the pole $S_1=s_{p_2}$ on the second Riemann sheet (resonance), and from the pole $s_1=s_{p_1}$ on the first Riemann sheet (bound state). (b) Numerically computed behavior of the survival probability $P_s(t)=|c_a(t)|^2$ versus normalized time $Jt$ for parameter values $g_0/J=1.2$ and $\gamma/J=2$. }
\end{figure}

We emphasize that, at such a critical value of the dissipation, neither the non-Hermitian Hamiltonian \(H_{\mathrm{NH}}\) nor the Liouvillian \(\mathcal{L}\) exhibit any exceptional point (EP) in their spectra. This is because the pole coalescence responsible for the transition occurs on the second Riemann sheet and therefore lies outside the spectrum of \(H_{\mathrm{NH}}\) (see Appendix~A). For this reason, we refer to it as a \emph{virtual exceptional point}.
Nevertheless, this virtual EP induces a qualitative change in the transient relaxation dynamics as \(\gamma\) is tuned across \(\gamma_{c_2}\), as illustrated in Figs.~7(b) and (d). For \(\gamma < \gamma_{c_2}\), the two poles on the second Riemann sheet share the same real part (decay rate) but have opposite imaginary parts [Fig.~7(a)]. Their interference produces an intermediate damped oscillatory behavior of \(c_a(t)\), with damping rate \(|\mathrm{Re}(s_{p_1})|=|\mathrm{Re}(s_{p_2})|>\gamma\) and oscillation frequency \(\Omega = 2|\mathrm{Im}(s_{p_1})|\). At longer times, this regime is supplanted by damped oscillations with a slower decay and a different period, originating from the contribution of the Hankel paths. The crossover between these two behaviors -- marked by an abrupt change in both decay rate and oscillation period -- occurs around \(t \simeq 7/J\), as shown in Fig.~7(b). Moreover, as the virtual EP is approached, the oscillation frequency \(\Omega\) in the initial transient regime gradually decreases and eventually vanishes. Above the virtual EP, i.e. for $\gamma_{c_2} < \gamma < \gamma_{c_1}$, the two poles are real, the decay behavior in the intermediate scale does not show oscillations anymore, and the relaxation dynamics is reshaped as illustrated in Fig.7(d).

\subsubsection{Strong-coupling regime}
The strong coupling regime corresponds to the condition $g_0 / J > \sqrt{2}$. A typical behavior of the real and imaginary parts of the two poles $s_{p_1}$ and $s_{p_2}$ versus normalized loss rate $\gamma /J$ is shown in Figs.9(a) and (b). As it can be seen, for $\gamma< \gamma_{c_1}=g_0^2/J$ the two poles $s_{p_1}$ and $s_{p_2}$ lie on the first Riemann sheet and correspond to two atom-photon bound states. Therefore,  the decay law (16) takes the form [Figs.10(a) and (c)]
\begin{equation}
c_a(t)=r_1 \exp(s_1 t)+r_2 \exp(s_2 t)+H_1(t)+H_2(t).
\end{equation}
\begin{figure}
\includegraphics[width=8.5 cm]{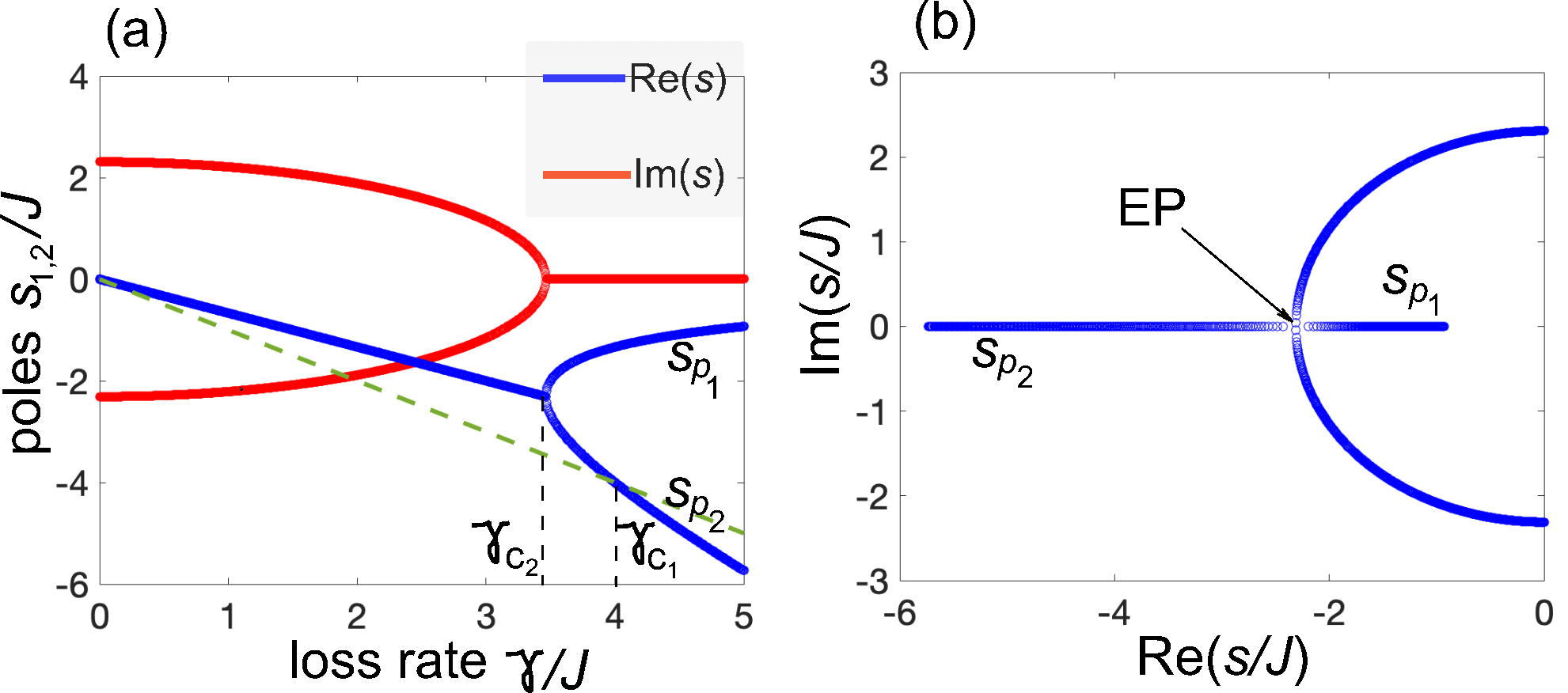}  
\caption{(a) Behavior of the real and imaginary parts of the two poles $s_{p_1}$ and $s_{p_2}$ versus loss rate $\gamma$ in the strong coupling regime ($g_0/J=2$). The dashed curve, corresponding to ${\rm Re}(s)=-\gamma$, separates the two Riemann sheets. Both poles contribute to the relaxation dynamics. For $\gamma< \gamma_{c_1}=g_0^2/J$, the two poles lie on the first Riemann sheet and correspond to two atom-photon bound states. At $\gamma=\gamma_{c_2}$, with $\gamma_{c_2}$ given by Eq.(27), the two poles coalesce, corresponding to an EP of the NH Hamiltonian. (b) Loci of the two poles in complex $s$ plane as the loss rate $\gamma$ is varied from $\gamma=0$ to $\gamma=5J$.}
\end{figure}
where $s_1=s_{p_1}$ and  $s_2=s_{p_2}$. Interestingly, a coalescence of the two atom-photon bound states occurs at $\gamma=\gamma_{c_2}$, where $\gamma_{c_2}$ is defined by Eq.(27). This corresponds to an EP of the NH Hamiltonian $H_{NH}$.
 The Hankel path contributions $H_{1,2}(t)$ are responsible for deviations from exponential decay at short time scale (Zeno regime). However, contrary to the weak- and moderate-coupling regimes, they do not dominate the asymptotic (long-time) relaxation dynamics, which is in fact established by the contributions of the two atom-photon bound states. For $\gamma< \gamma_{c_2}$, the two poles are complex conjugate, with the same decay rate and opposite imaginary parts, resulting in an asymptotic damped oscillatory relaxation at the frequency $\Omega=2 |{\rm Im}(s_{p_1})|$ and damping rate $|{\rm Re}(s_{p_1})| < \gamma$ [Fig.10(b)]. On the other hand, for $\gamma_{c_2} < \gamma < \gamma_{c_1}$ the relaxation dynamics is purely exponential and dominated by the pole $s_1$ [Fig.10(d)]. Interestingly, an inspection of Fig.9 indicates that the fastest relaxation of the quantum emitter occurs at the loss rate $\gamma=\gamma_{c_2}$, i.e. at the EP. This is a distinctive feature than the moderate-coupling regime, where the fastest relaxation of the emitter occurs at $\gamma=\gamma_{c_1}$.\\
 Finally, for $\gamma> \gamma_{c_1}$ one of the two poles, $s_{p_1}$,  remains localized on the first Riemann sheet and corresponds to an atom-photon bound state, whereas the other pole, $s_{p_2}$, crosses the branch cut and is located on the second Riemann sheet (resonance); see Fig.11(a).  The decay law (16) then takes the form 
 \begin{equation}
c_a(t)=r_1 \exp(s_1 t)+R_1 \exp(S_1 t)+H_1(t)+H_2(t).
\end{equation}
where $s_1=s_{p_1}$ and $S_1=s_{p_2}$. The intermediate and long-time relaxation dynamics is exponential and determined by the $s_1$ pole contribution, as shown in Fig.11(b).
\par

\begin{figure}
\includegraphics[width=8.5 cm]{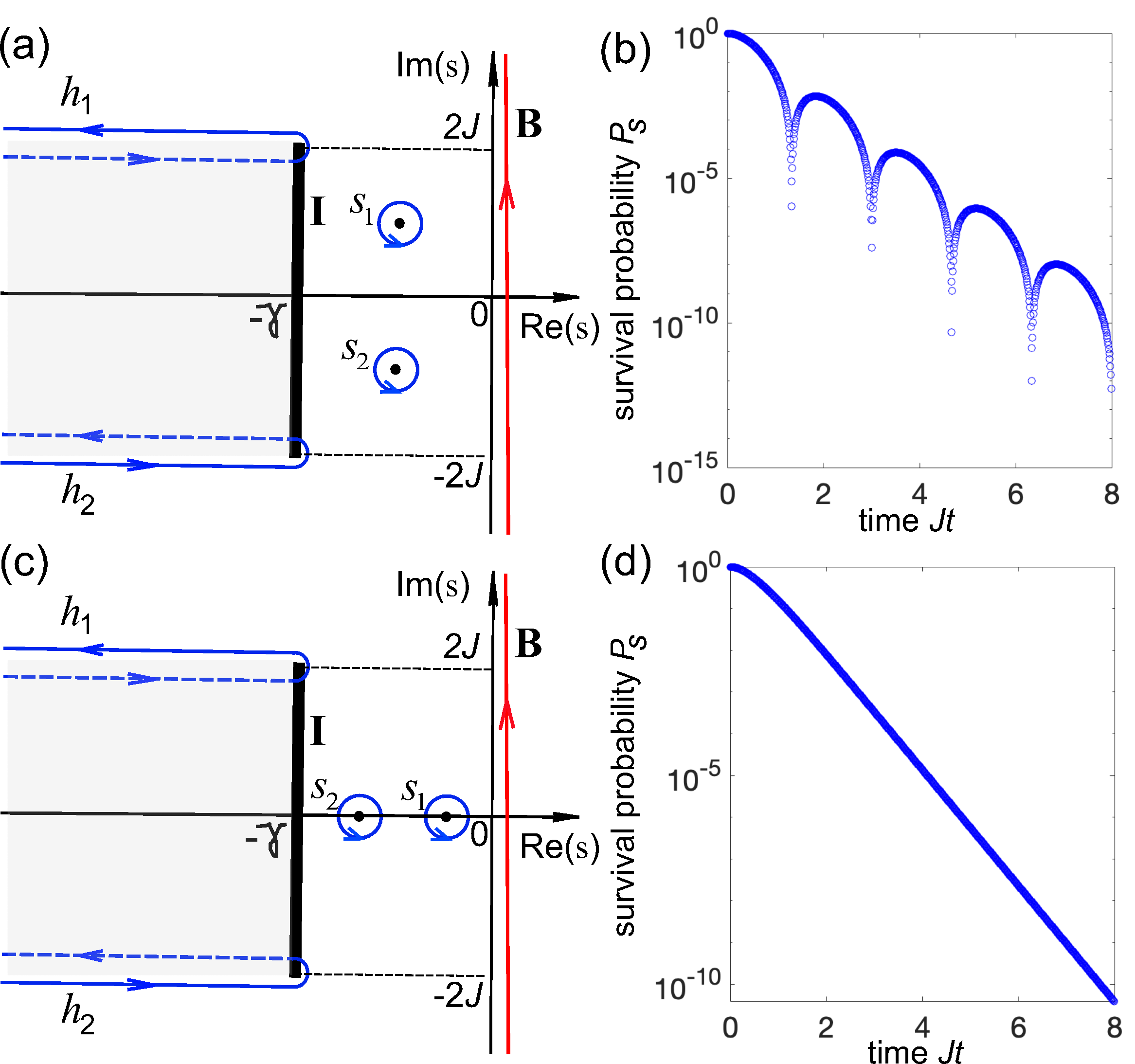}  
\caption{(a,c) Integration contour paths in complex $s$ plane in the strong-coupling regime for (a) $\gamma< \gamma_{c_2}$, and (c) $\gamma_{c_2}<\gamma<\gamma_{c_1}$. The contribution to the decay law $c_a(t)$ comes from the two Hankel paths $h_{1,2}$ and from the two poles $s_1=s_{p_1}$ and $s_2=s_{p_2}$  on the first Riemann sheet. The asymptotic long-time relaxation dynamics is dominated by the two poles. An EP, corresponding to the coalescence of the two poles on the first Riemann sheet (atom-photon bound states), occurs at $\gamma=\gamma_{c_2}$. (b,d) Numerically computed behavior of the survival probability $P_s(t)=|c_a(t)|^2$ versus normalized time $Jt$ for $g_0/J=2$. In (b),  $\gamma/J=2$, corresponding to $\gamma< \gamma_{c_2}$; in (d), $\gamma /J=3.7$, corresponding to $\gamma_{c_2}<\gamma<\gamma_{c_1}$. Note the different long-time relaxation dynamics below and above the EP.}
\end{figure}

\begin{figure}
\includegraphics[width=8.5 cm]{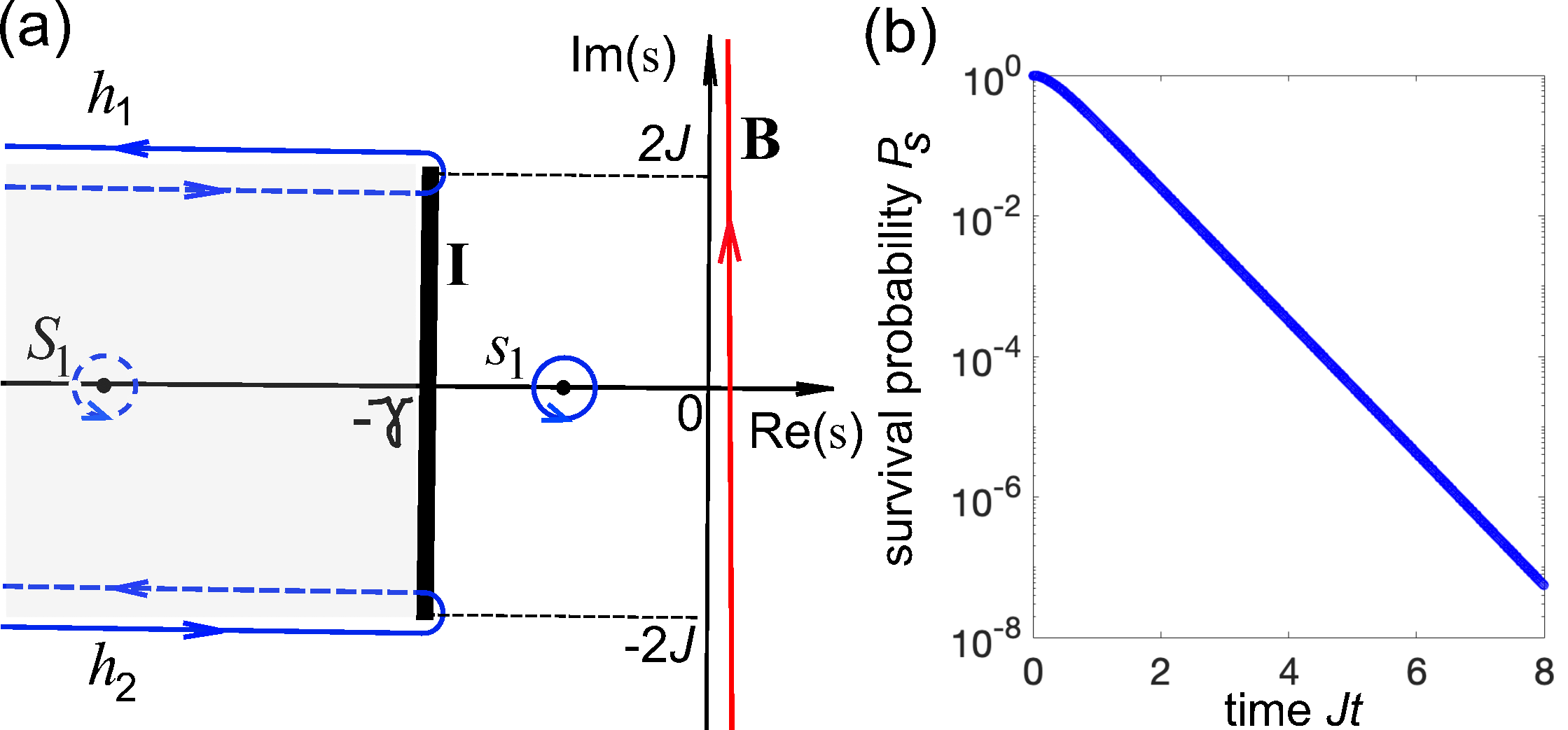}  
\caption{(a) Integration contour in complex $s$ plane in the strong-coupling regime and for $\gamma> \gamma_{c_1}$. The contribution to the decay law $c_a(t)$ comes from the two Hankel paths $h_{1,2}$, from the pole $S_1=s_{p_2}$ on the second Riemann sheet (resonance), and from the pole $s_1=s_{p_1}$ on the first Riemann sheet (bound state). The intermediate and long-time relaxation dynamics is dominated by the pole $s_1$ contribution. (b) Numerically computed behavior of the survival probability $P_s(t)=|c_a(t)|^2$ versus normalized time $Jt$ for parameter values $g_0/J=2$ and $\gamma/J=4.5$. }
\end{figure}

The above analysis indicates that dynamical behavior of the system arises from the combined contributions of bound states, resonant states, and the Hankel path integrals (continuum). 
From an experimental perspective, the distinct dynamical contributions could, in principle, be resolved through time-resolved measurements of the emitter survival probability. Bound states manifest as an asymptotic exponential decay that dominates at long evolution times, reflecting the discrete poles of the resolvent on the physical Riemann sheet. In contrast, resonant states give rise to transient, exponentially damped oscillations at intermediate times, whose decay rates are governed by the imaginary parts of the resonance poles located on the second Riemann sheet. The remaining contribution, originating from the Hankel-path integrals along the branch cut, dominates the early-time dynamics and produces non-exponential, algebraic decay tails. These regimes could be probed experimentally in ultrafast waveguide-QED or photonic-lattice platforms capable of tracking emitter dynamics across both Zeno and asymptotic timescales.

\section{Conclusion and Discussion}
In this work, we investigated the relaxation dynamics of a quantum emitter coupled to a semi-infinite uniformly lossy lattice, a minimal yet physically relevant model in waveguide QED. While non-Hermitian baths have been considered in previous studies, our main novelty is to unveil the \emph{coexistence} of ordinary and virtual exceptional points (EPs) in such a dissipative environment, and to demonstrate that these distinct singularities can drive dynamical phase transitions in the spontaneous-emission decay of the emitter. Despite the apparent simplicity of the model, the interplay between ordinary EPs (discrete eigenvalue/eigenvector coalescences of effective non-Hermitian descriptions) and virtual EPs (coalescence of resonance poles on the second Riemann sheet) produces a rich dynamical landscape as the bath dissipation is varied.

A central result is the identification of an intermediate regime of optimal dissipation, where the emitter relaxes faster than in either the weak- or strong-loss limits. This shows that dissipation can be harnessed as a controllable resource to accelerate relaxation and enhance light-matter coupling. The microscopic origin of these effects is spectral: phase transitions in the decay dynamics are traced to a restructuring of the resolvent poles on the first and second Riemann sheets, and in particular to the coalescence of resonant states on the second sheet. These resonance coalescences manifest in transient relaxation and are naturally interpreted as virtual EPs, highlighting both their conceptual link to conventional EP physics and their distinctive character in infinite-dimensional NH systems.
More broadly, our results underscore that the \emph{nature} of the dissipative bath -- not merely the presence of loss -- crucially determines emitter dynamics. In the uniformly lossy lattice studied here, the dynamics are governed by the spectral properties of the effective non-Hermitian Hamiltonian, with resonance and/or bound state coalescence playing the dominant role and quantum-jump processes remaining irrelevant. By contrast, when losses are applied only to alternating lattice sites, the relaxation can display critical (algebraic) decay \cite{T18,T19}, a form of temporal criticality that parallels the power-law spatial decay of correlations in conventional critical phases. An equally distinct scenario arises in baths subject to local dephasing rather than particle loss. In this case, particle number is conserved and quantum jumps play a central role in the relaxation dynamics, which cannot be fully captured by an effective non-Hermitian description; correspondingly, the decay follows a critical algebraic law \cite{unpublished}. These examples illustrate how different types of dissipation -- uniform versus staggered, loss versus dephasing -- can profoundly alter both the qualitative and quantitative features of spontaneous emission, determining whether relaxation is exponential or algebraic and whether its behavior is governed by spectral pole coalescences or stochastic quantum jumps.

Uniformly lossy lattices are therefore both analytically tractable and experimentally accessible testbeds for exploring the coexistence and interplay of ordinary and virtual EPs and their impact on dynamical phase transitions in waveguide QED. Natural extensions include multi-emitter configurations, higher-dimensional baths, and explicitly non-Markovian environments; each avenue may reveal new mechanisms by which different dissipation types alter EP formation, spectral restructuring, and relaxation. In multi-emitter setups, both bound and resonant states can mediate interactions, with resonances giving rise to transient, time-dependent couplings that decay over the resonance lifetime. Such effects may lead to temporally modulated, non-Markovian dynamics and constitute an intriguing direction for future studies.

More broadly, these findings point toward dissipation engineering as a versatile tool for quantum technologies: by tailoring the \emph{type} and spatial structure of losses or dephasing, one can design environments that either protect quantum coherence or accelerate relaxation as required by the application.
%\section*{Supporting Information}
%Supporting Information is available from the Wiley Online Library or from
%the author.

\section*{Acknowledgements}
The author acknowledges the
Spanish Agencia Estatal de Investigacion (MDM-2017-0711).

\section*{Conflict of Interest}
The author declares no conflict of interest.

\section*{Data Availability Statement}
The data that support the findings of this study are available from the corresponding
author upon reasonable request.

\section*{Keywords}
non-Hermitian physics, waveguide QED, phase transitions, exceptional points, quantum mechanical decay

\appendix

\section{Physical and virtual exceptional points}

In this Appendix, we clarify the distinction between ordinary (or physical) EPs and virtual EPs. Let us consider an effective non-Hermitian Hamiltonian $H_{NH}$, describing the system dynamics, which depends on a set of control parameters $\lambda$. The spectrum of $H_{NH}$ is entirely determined by the singularities of the resolvent  
\begin{equation}
G(z) = \frac{1}{z - H_{NH}}
\end{equation}  
in the complex energy plane. The nature of the singularities of the resolvent depends crucially on the dimensionality of the system, i.e. whether $H_{NH}$ acts on a finite- or infinite-dimensional Hilbert space.  

In the finite-dimensional case, $H_{NH}$ is represented by an $N \times N$ matrix, and the singularities of $G(z)$ are given by its $N$ poles $z_1(\lambda), z_2(\lambda), \ldots, z_N(\lambda)$, which coincide with the complex eigenvalues of the matrix $H_{NH}$. As the control parameters $\lambda$ are varied, an ordinary (or physical) EP occurs at a critical value $\lambda = \lambda_c$ when two (or more) eigenvalues coalesce, along with their corresponding eigenvectors. At $\lambda=\lambda_c$, the matrix $H_{NH}$ becomes defective and cannot be diagonalized.  

In infinite-dimensional NH systems, the structure of the resolvent is richer. In general, $G(z)$ can display two distinct types of singularities (we do not consider here a third type, spectral singularities, which can arise in the continuous spectrum of $H_{NH}$ and have been discussed in some previous works \cite{SS1,SS2,SS3,SS4}). First, if the point spectrum of $H_{NH}$ is not empty and contains a set of discrete complex energies $z_1(\lambda), z_2(\lambda), \ldots$, then $G(z)$ exhibits poles at $z=z_j(\lambda)$, whose corresponding eigenvectors are normalizable bound states. As in the finite-dimensional case, an ordinary EP can arise when two (or more) of these poles coalesce at a critical parameter value $\lambda=\lambda_c$, along with their associated bound eigenstates.  
Second, in infinite-dimensional systems the resolvent can also develop branch cuts along curves I in the complex energy plane, across which $G(z)$ is discontinuous. The complex energies on these branch cuts correspond to the absolutely continuous spectrum of $H_{NH}$, with associated non-normalizable scattering eigenstates. For example, in the waveguide QED model discussed in the main text, the resolvent element  
\begin{equation}
G_{e,e}(z) = \langle e | G(z) | e \rangle
\end{equation}  
is related to the Laplace transform $\hat{c}_{a}(s)$ of the excited-state amplitude $c_a(t)$ [Eq.(13)] via  
\begin{equation}
G_{e,e}(z) = -i \hat{c}_a(s=-iz).  
\end{equation}  
Explicitly,  
\begin{equation}
G_{e,e}(z) = \frac{1}{z - \Delta \omega_0 - \frac{g_0^2}{J^2}\left( z+i\gamma - \sqrt{(z+i\gamma)^2 - 4J^2} \right)}.
\end{equation}  
The square root in the denominator of Eq.(A4) introduces a branch cut along the segment ${\rm I} = (-2J-i\gamma,\, 2J-i\gamma)$ on the line ${\rm Im}(z)=-\gamma$, corresponding to the cut shown in Fig.2 after the substitution $s=-iz$.  

Because $G(z)$ is discontinuous across the branch cut, $G(z)$ can be analytically continued to a second Riemann sheet, when crossing the cut, from one side to the other one, yielding $G^{(II)}(z)$. The analytic continuation should be considered when deforming the path integral in Eq.(13) to cross the branch cut, as shown in Fig.2. The poles of $G^{(II)}(z)$, denoted by $Z_1(\lambda), Z_2(\lambda), \ldots$, are distinct from those on the first sheet. These poles do not belong to the spectrum of $H_{NH}$; instead, they correspond to ``unphysical'' solutions of the Schr\"odinger equation $H_{NH}|\psi\rangle = z|\psi\rangle$, whose eigenstates -- known as Gamov' or Siegert's states \cite{GA-1,GA4b}-- are neither normalizable bound states nor scattering states. Such states are typically classified as resonances, anti-resonances, and anti-bound states \cite{GA1b,GA2,GA4}; for the specific waveguide QED model  considered in the main text they are discussed in Appendix B. Despite their unphysical nature, resonances play a central role in shaping transient dynamics and scattering features such as Breit--Wigner or Fano resonances \cite{GA-1,GA0,GA0b}.  

A \emph{virtual EP} is then defined as the coalescence of two (or more) resonance poles $Z_k(\lambda)$ of the analytically continued resolvent on the second Riemann sheet, together with their associated Siegert's eigenstates, at a critical parameter value $\lambda=\Lambda_c$. Importantly, at $\lambda=\Lambda_c$ the Hamiltonian $H_{NH}$ remains diagonalizable and does not host any EP in its physical spectrum---hence the terminology ``virtual'' EP.  

\begin{figure}
\includegraphics[width=6 cm]{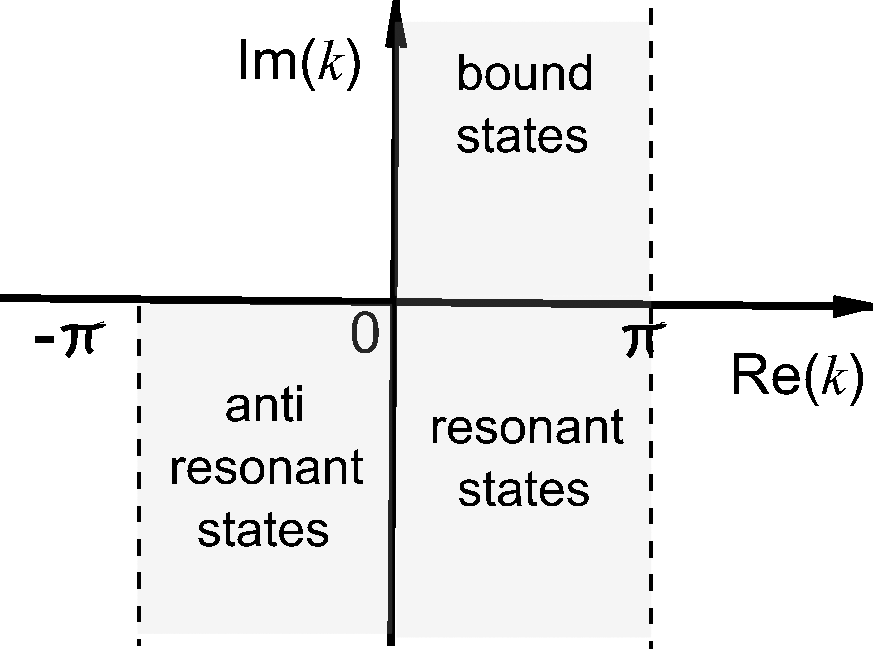}  
\caption{Typology of eigenstates of the equation $H_{NH} | \psi \rangle=z | \psi \rangle$  in the complex $k$ plane under Siegert boundary conditions, illustrating resonant, anti-resonant and bound states.}
\end{figure}
\begin{figure*}
\includegraphics[width=16 cm]{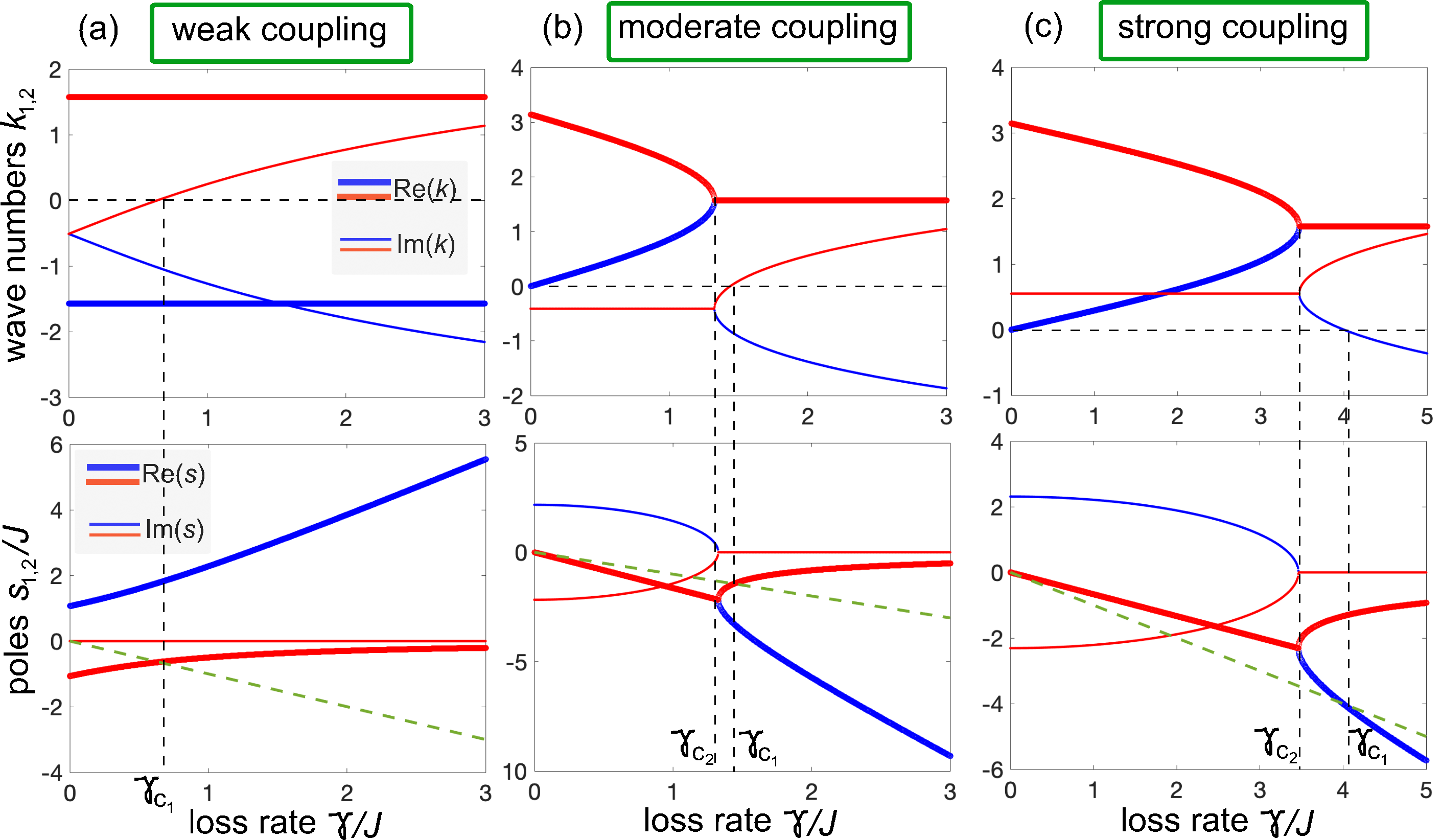}  
\caption{Typical behavior of the complex Bloch wave numbers $k_{1,2}$ (upper plots) and corresponding complex energies $z=is$ (lower plots) of bound and resonant/anti-resonant states, as given by Eqs.(B4) and (B5), versus loss rate $\gamma/J$ in the three coupling regimes. (a) Weak-coupling regime ($g_0 /J=0.8$), (b) moderate-coupling regime ($ g_0/J=1.2$), and (c) strong-coupling regime ($g_0/J=2$). The bold (thin) curves refer to the real (imaginary) parts of $k=k_R+ik_I$ and $s$, whereas the two colors (red and blue) refer to the two roots. The green dashed curve in the lower panels is the normalized loss rate $\gamma/J$.  States with $k_I>0$ are bound states, whereas states with $k_I<0$ are resonant states for $k_R>0$ or anti-resonant states for $k_R<0$. The real part of $s$ for a resonant state is alway smaller than $-\gamma$, i.e. it is located below the green dashed curves. The critical values $\gamma_{c_1}$ and $\gamma_{c_2}< \gamma_{c_1}$ of dissipation rates are given by Eqs.(22) and (27) in the main text. In the weak-coupling regime [panels (a)], there is one resonant and one anti-resonant state for $\gamma<\gamma_{c_1}$, while there is a bound state and a resonant state for $\gamma> \gamma_{c_1}$. In the moderate-coupling regime [panels (b)] there are two resonant states for  $\gamma<\gamma_{c_1}$, while there is a bound state and a resonant state for $\gamma> \gamma_{c_1}$. At  $\gamma=\gamma_{c_2}$ the two resonant states coalesce, corresponding to a virtual EP. Finally, in the strong-coupling regime [panels (c)] there are two bound states for  $\gamma<\gamma_{c_1}$, while there is a bound state and a resonant state for $\gamma> \gamma_{c_1}$. At  $\gamma=\gamma_{c_2}$ the two bound states coalesce, corresponding to a physical EP. }
\end{figure*}

\section{Atom-photon bound states and resonant states}
Bound atom-photon states and resonant states of the effective NH Hamiltonian $H_{NH}$ can be determined by solving the eigenvalue problem 
\begin{equation}
H_{NH} | \psi \rangle=z | \psi \rangle
\end{equation}
 with Siegert boundary conditions \cite{GA-1,GA1b,GA4b,D5b}, where $z$ is the complex energy of the state related to the pole $s$ of $\hat{c}_a(s)$ via the simple relation $z=is$.  From Eq.(7), the eigenvalue problem (B1) takes the form
 \begin{eqnarray}
z c_a & = & \Delta \omega_0 c_a + g_0 b_1, \nonumber \\
z b_1  & = & g_0 c_a - J b_2 - i \gamma b_1, \\
z b_n & = & -J (b_{n+1} + b_{n-1}) - i \gamma b_n, \quad n \ge 2. \nonumber
\end{eqnarray}
  The Siegert boundary conditions correspond to assuming a single plane wave, with {\em complex} Bloch wave number $k$, in the semi-infinite lattice, i.e. by letting in Eq.(B2)
  \begin{equation}
  b_n=\exp(ik n)=X^n
  \end{equation}
where $X=\exp(ik)$ and where the real part of $k$, $k_R$, is assumed to vary in the range $-\pi < k_R < \pi$. The Siegert boundary condition is basically equivalent to state that, in a scattering problem on the semi-infinite lattice  with an edge defect, described by $H_{NH}$ [Eq.(B2)], the reflection amplitude vanishes or diverges for an incident wave.
 Substitution of Eq.(B3) into Eq.(B2) and eliminating the amplitude $c_a$ yields
\begin{equation}
k=-i \log X=-i \log \left\{  \frac{1}{J} \left(  \frac{g_0^2}{z-\Delta \omega_0} -z-i \gamma \right)\right\}
\end{equation}
where $z$ in a root of the second-order algebraic equation
\begin{eqnarray}
(1-2 \sigma) z^2-2iz \left\{ \sigma \gamma +i (\sigma-1) \Delta \omega_0  \right\}  \nonumber \\ 
+4 \sigma^2 J^2 +\Delta \omega_0^2+2i \gamma \sigma \Delta \omega_0=0
\end{eqnarray}
with $\sigma \equiv g_0^2/(2J^2)$. Note that, after letting $z=is$, Eq.(B5) is precisely Eq.(20) given in the main text that determines the poles of $\hat{c}_a(s)$ on the first or second Riemann sheets. Since Eq.(B5) is of second order, there are two roots $z_{1,2}$, i.e. $s_{1,2}=-i z_{1,2}$, with corresponding complex wave numbers $k_{1,2}$ obtained from Eq.(B4). 
An atom-photon bound state, corresponding to an eigenenergy $z$ belonging to the point spectrum of $H_{NH}$, is a root to Eq.(B5) corresponding to a positive imaginary part of $k$, i.e. $k_I>0$ or equivalently $|X|<1$. It can be readily shown that for an atom-photon bound state one has $0 \geq {\rm Re}(s) \geq  -\gamma$, i.e. the decay rate of the atom-photon bound state is smaller than the loss rate $\gamma$ of the bath.  Conversely, a root to Eq.(B5) with a negative imaginary part of $k$, i.e. $k_I<0$ or equivalently $|X|>1$, is unbounded as $n \rightarrow \infty$ and thus does not belong to the point spectrum nor to the absolutely continuous spectrum of $H_{NH}$. Such "unphysical" states correspond to resonant states for $0< k_R<\pi$ (outgoing waves) and to anti-resonant states for $-\pi < k_R<0$ (incoming waves) \cite{GA1b}; see Fig.12 for a schematic. It can be readily shown that for a resonant state one has ${\rm Re}(s)< -\gamma$, i.e. the decay rate of a resonant state is always larger than the dissipation rate $\gamma$ of the bath, whereas for anti-resonant states one has ${\rm Re}(s)>0$, so that they do not contribute to the spontaneous emission process of the quantum emitter. The existence of bound, resonant and anti-resonant states is greatly affected by the atom-photon coupling rate $g_0/J$ and loss rate $ \gamma/J$ of the bath. Typical behaviors of real and imaginary parts of the complex Bloch wave number $k=k_R+ik_I$ and of the two roots $s_{1,2}=-iz_{1,2}$ of the algebraic equation (20) versus loss rate $\gamma /J$ in the three coupling regimes (weak-coupling $g_0/J<1$, moderate-coupling $1<g_0/J< \sqrt{2}$ and strong-coupling $g_0/J> \sqrt{2}$) are illustrated in Fig.13. 
Note that in the weak-coupling regime [Fig.13(a)], there is one resonant state and one anti-resonant state for $\gamma<\gamma_{c_1}$, while there is a bound state and a resonant state for $\gamma> \gamma_{c_1}$. In the moderate-coupling regime [Fig.13(b)] there are two resonant states for  $\gamma<\gamma_{c_1}$, while there is a bound state and a resonant state for $\gamma> \gamma_{c_1}$. At  $\gamma=\gamma_{c_2}$ the two resonant states coalesce, corresponding to a virtual EP. Finally, in the strong-coupling regime [Fig.13(c)] there are two bound states for  $\gamma<\gamma_{c_1}$, while there is a bound state and a resonant state for $\gamma> \gamma_{c_1}$. At  $\gamma=\gamma_{c_2}$ the two bound states coalesce, corresponding to an EP of the NH Hamiltonian.

\end{document}